\documentclass[floatfix,amsmath,amssymb,superscriptaddress,longbibliography,twocolumn,showpacs]{revtex4-1}
\usepackage{amssymb}
\usepackage{amsmath}
\usepackage{bbold}
\usepackage{bm}

\usepackage{braket}
\usepackage[title]{appendix}
\usepackage{xcolor}
\usepackage{soul}
\usepackage{graphicx}
\usepackage{textcomp}
\usepackage{enumerate}
\usepackage{multirow}
\usepackage{hyperref}
\usepackage{lipsum, babel}
\graphicspath{{../figura1/}{../figura2/}}

\newcommand*\diff{\mathop{}\!\mathrm{d}}
\newcommand*\Tr{\mathop{}\!\mathrm{Tr}}

\usepackage{soul}

\newcommand{\reference}[1]{Ref.~\cite{#1}}

\makeatletter
\renewcommand{\paragraph}[1]{%
  \textit{#1}\hspace{0.5em}%
}
\makeatother

\begin{document}

\title{Kernel shape renormalization explains output-output correlations in finite Bayesian one-hidden-layer networks}

\author{P. Baglioni}
\affiliation{INFN, sezione di Milano Bicocca, Piazza della scienza 3, 20126, Milano, Italy}
\author{L. Giambagli}
\affiliation{Department of Physics and Astronomy, University of Florence, INFN and CSDC, 50019 Sesto Fiorentino, Italy}
\author{A. Vezzani}
\affiliation{Istituto dei Materiali per l'Elettronica ed il Magnetismo (IMEM-CNR), Parco Area delle Scienze, 37/A-43124 Parma, Italy}
\affiliation{Dipartimento di Scienze Matematiche, Fisiche e Informatiche,
Universit\`a degli Studi di Parma, Parco Area delle Scienze, 7/A 43124 Parma, Italy}
\affiliation{INFN, Gruppo Collegato di Parma, Parco Area delle Scienze 7/A, 43124 Parma, Italy}
\author{R. Burioni}
\affiliation{Dipartimento di Scienze Matematiche, Fisiche e Informatiche,
Universit\`a degli Studi di Parma, Parco Area delle Scienze, 7/A 43124 Parma, Italy}
\affiliation{INFN, Gruppo Collegato di Parma, Parco Area delle Scienze 7/A, 43124 Parma, Italy}
\author{P. Rotondo}
\affiliation{Dipartimento di Scienze Matematiche, Fisiche e Informatiche,
Universit\`a degli Studi di Parma, Parco Area delle Scienze, 7/A 43124 Parma, Italy}
\affiliation{INFN, Gruppo Collegato di Parma, Parco Area delle Scienze 7/A, 43124 Parma, Italy}
\author{R. Pacelli}
\affiliation{INFN, sezione di Padova, Via Marzolo 8, 35131 Padova, Italy}

\begin{abstract}
Finite-width one hidden layer networks with multiple neurons in the readout layer display non-trivial output-output correlations that vanish in the lazy-training infinite-width limit.
In this manuscript we leverage recent progress in the proportional limit of Bayesian deep learning (that is the limit where the size of the training set $P$ and the width of the hidden layers $N$ are taken to infinity keeping their ratio $\alpha = P/N$ finite) to rationalize this empirical evidence.
In particular, we show that output-output correlations in finite fully-connected networks are taken into account by a kernel shape renormalization of the infinite-width NNGP kernel, which naturally arises in the proportional limit.
We perform accurate numerical experiments both to assess the predictive power of the Bayesian framework in terms of generalization, and to quantify output-output correlations in finite-width networks. By quantitatively matching our predictions with the observed correlations, we provide additional evidence that kernel shape renormalization is instrumental to explain the phenomenology observed in finite Bayesian one hidden layer networks.
\end{abstract}

\maketitle
\paragraph{Introduction---}
Delving into the mechanics of deep learning is considered a crucial challenge for theorists. Given the complex nature of this problem, only a handful of analytically tractable models are available.
The first successful attempt to investigate a two-layer (equivalently, one hidden layer, 1HL) Neural Network (NN), by R. Neal \cite{Neal}, relies on the assumption that the width of the hidden layer $N_1$ is large compared to the train set size $P$. This large-width limit, later known as lazy-training infinite-width, was extensively explored in subsequent literature \cite{mackay1998introduction, NIPS1996_ae5e3ce4,g.2018gaussian,LeeGaussian, aroracnn2019, ChizatLazy,aitchison2020bigger,favaro2023quantitative} and eventually extended to any feedforward architecture \cite{garriga-alonso2018deep,novak2019bayesian,tensori}. 

Nevertheless, Neal himself noticed that such infinite networks with $D$ units in the readout layer end up to be equivalent to $D$ separate networks with a single output \cite{Neal}. This occurs because output functions are composed of numerous contributions from hidden units, where each individual contribution is negligible in the large $N_1$ limit. As a result, hidden units do not represent “hidden features" that capture important aspects of
the data. For the network to do this, there needs to be connection among the outputs, since they come from shared hidden representations. Analysing this point is crucial, since it is widely believed that the success of deep learning lies in the ability of NNs to perform meaningful feature extraction \cite{yu2013feature,petrini2022learning, Ciceri2024, shi2024spring}. Already in the nineties, D. Mackay pointed out that one major point of failure of the infinite-width description was indeed the inability to account for output-output correlations  observed in finite networks \cite{mackay1998introduction}.

Recently, progress has been made in the context of infinite-width networks, unveiling a feature-learning phase - as opposed to lazy-training - with the so-called mean-field scaling \cite{doi:10.1073/pnas.1806579115,NEURIPS2018_a1afc58c,doi:10.1137/18M1192184,NEURIPS2018_196f5641,tensoriv, cagnetta2023what, favero2021locality,yang2023theory}. The unrealistic assumption of the infinite-width limit is albeit in mismatch with real-world machine learning, where usually the number of samples exceeds the widths of the hidden layers. 

Another possibility to analytically inspect deep networks is working in the proportional thermodynamic regime, formally defined as the asymptotic limit where the width of the layers $N_\ell$ ($\ell = 1 \ldots L$, with $L+1$ being the finite network's depth) scales to infinity together with $P$, with their ratios $\alpha_\ell = P/N_\ell$ fixed \cite{SompolinskyLinear, pacelli2023statistical, aiudi2023, ingrosso2024tl, baglioni2024predictive, fischer24critical, cui2024highdim, camilli2023fundamental, tiberi2024dissecting}. As shown for a variety of different architectures, the proportional assumption allows extracting predictions for practically-relevant scores as observables over an equilibrium ensemble \cite{SompolinskyLinear, pacelli2023statistical, aiudi2023, ingrosso2024tl}, and test them on finite networks of arbitrary (but large) width $N$ \cite{baglioni2024predictive}. 

In the following, we leverage the aforementioned results to study two-layer fully-connected (FC) Bayesian networks with multiple outputs in the proportional regime. Such exploration yields (i) closed-form expression for the generalization loss after training, which we test on finite Bayesian networks (ii) a quantifiable measure of output-output correlations in terms of correlations between weights associated with different output neurons. The emergence of such correlations stems from the particular form of kernel renormalization that occurs in multiple-output FC networks, called kernel \textit{shape} renormalization, which is intrinsically different from that of single-output architectures \cite{SompolinskyLinear, pacelli2023statistical}. 
Finally, we compare the generalization performance of Bayesian models to that of 1HL  networks trained with a state-of-the-art (SOTA) algorithm, Adam \cite{adam}, and hyperparameter optimization strategy \cite{liaw2018tune}. We show that Adam is not superior in terms of performance on FC models for two computer-vision datasets, indicating that the solutions it finds are not qualitatively different from those in the Gibbs equilibrium ensemble. 

\paragraph{Problem setting---}
We analyze the supervised regression problem of learning train set $X= \{ x^\mu \in \mathbb R^{N_0} \}_{\mu=1}^P$, with corresponding $D$-dimensional labels $Y= \{ y^\mu \in \mathbb R^D \}_{\mu=1}^P$. 
For a FC 1HL network parametrized by the weights $\theta = \lbrace w,v \rbrace$ that implements the function $f(x;\theta) = (f_{a}, a = 1 \ldots D)$, we define the Mean Squared Error (MSE) loss as $\mathcal L =  \sum_{\mu=1}^{P} \Vert y^\mu - f(x^\mu;\theta) \Vert^2/2$, where $\Vert \cdot \Vert$ is the standard Frobenius norm. The weights of the first and last layer, respectively $w \in \mathbb R^{N_1 \times N_0}$ and $v \in \mathbb R^{D \times N_1}$, are taken with Gaussian prior distributions $w \sim  \mathcal{N}(\mathbf 0,  \mathbb 1/\lambda_0)$, $v \sim  \mathcal{N}(\mathbf 0,  \mathbb 1/ \lambda_1)$. 
The parameters $\lambda_0, \lambda_1$ can be equivalently thought as Gaussian priors over the weights of each layer, or as $L^2$ regularization terms rescaled by the temperature $T$. Denoting $\sigma$ a point-wise non-linear activation function, the network's outputs read $f(x) = v\sigma\left[h(x)\right]/\sqrt{N_1}$, with pre-activations $h(x) = wx/\sqrt{N_0}$.
\begin{figure}[!hb]
\centering
\includegraphics[width=1.\linewidth]{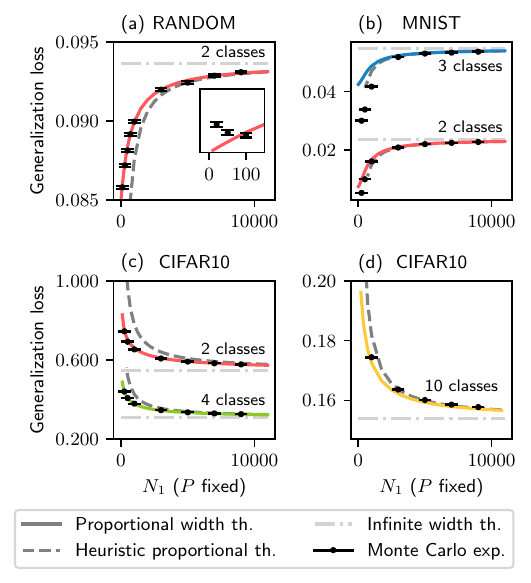}
\caption{\textbf{Generalization performance of finite Bayesian shallow networks with multiple outputs.} Generalization loss of a 1HL network with $D$ outputs as a function of the hidden layer size $N_1$ for different datasets. In all plots, solid lines represent predictions obtained with $Q^*$ that minimizes Eq. \eqref{eq:S_FC}, dashed lines are predictions obtained with a heuristic solution for the order parameters (see main text), dot-dashed lines indicate the infinite-width prediction, and points with error bars represent numerical experiments within one standard deviation from the mean. (a) Random Gaussian data with labels given by a linear teacher (see SM for more details) with $N_0 = 12^2$ and $ D = 2$ classes. (b) MNIST dataset ($N_0 = 28^2$) with $D = 2,3$, respectively using examples with labels (“$0$", “$1$") and (“$1$",“$2$",“$3$"). (c-d) CIFAR10 dataset downscaled to $N_0 = 28^2$, with $D = 2$ (“cats", “horses"), $D = 4$ (“cars", “birds", “cats", “deers"), and $D=10$ (all classes). In all plots, we use $ P= P_{\mathrm{test}} = 1000 $, $\lambda_0 = \lambda_1 = 1.0$, $T=0.01$ and set the biases to zero. Both for MNIST and CIFAR10, we perform one-hot-encoding of classes, so each label is a $D$-dimensional vector with zeros everywhere except for one entry (whose position determines the class of the example).
}
\label{fig:1}
\end{figure}

\paragraph{Effective action for shallow FC networks---} 
The canonical partition function at temperature $T=1/\beta$ associated to the train loss function $\mathcal L$ reads $Z = \int \diff \mu (\theta)  \, \exp \{-\beta \mathcal{L}(\theta)  \}$, where $\diff \mu(\theta)$ indicates collective integration over the weights and their prior distribution $\mu(\theta)$. In the thermodynamic proportional limit $ N_1, P \to \infty$ at fixed ratio $\alpha_1 = P/N_1$, it is possible to reduce $Z$ to a form suitable to saddle-point integration over an ensemble of $D \times D$ positive-definite matrix order parameters $ Q$, $Z = \int d Q \, \mathrm{exp}\{-N_1 S_{\textrm{FC}}(Q)/2\}$. Denoting $\mathbf y$ the vectorized matrix $Y$, and restricting to odd activation functions, the action is given by \cite{pacelli2023statistical}:
\begin{align}
    S_{\textrm{FC}}&=  \Tr Q  - \log\det Q +\frac{\alpha_1}{P} \mathbf y^T \left(  \frac{\mathbb{1}}{\beta} + K_Q^{(\mathrm{R})}(X,X)\right)^{-1} \mathbf y\nonumber \\
    &+\frac{\alpha_1}{P}\text{Tr}\log \beta \left( \frac{\mathbb{1}}{\beta} +K_Q^{(\mathrm{R})}(X,X)\right) \,.
    \label{eq:S_FC}
\end{align}
The renormalized kernel $K_Q^{(\mathrm{R})}(X,X)$ is a $PD \times PD$ matrix defined as: 
\begin{equation}
    K_Q^{\mathrm{(R)}}(X,X) = Q \otimes K^{\mathrm{NNGP}} (X,X) \, ,
    \label{eq:renorm_kernel}
\end{equation}
where $\otimes$ denotes the outer product, and $K^{\mathrm{NNGP}} (X,X)$ is the Neural Network Gaussian Process (NNGP) kernel function $K^{\mathrm{NNGP}} (x,x')$  evaluated for pairs of train inputs $[K^{\mathrm{NNGP}}(X,X)]_{\mu\nu}=K^{\mathrm{NNGP}} (x^\mu,x^\nu)$ \cite{LeeGaussian}. 
Minimization of Eq. \eqref{eq:S_FC} is straightforward if $\alpha_1 \to 0$, yielding $Q^* = \mathbb 1$ and recovering the infinite-width limit result \cite{LeeGaussian}. At finite $\alpha_1$, one finds data-dependent solutions for the matrix order parameter that produce a shape renormalization of the infinite-width kernel, as expressed by Eq. \eqref{eq:renorm_kernel}. Note that, in the single-output case $D=1$, the matrix $Q$ reduces to a scalar, and the corresponding kernel renormalization does not change the NNGP kernel's shape.\cite{pacelli2023statistical}.

The effective action at proportional width presented in Eq. \eqref{eq:S_FC} has been firstly derived for deep linear networks in Ref. \cite{SompolinskyLinear}. The non-asymptotic evaluation of the partition function in the linear case is given in Ref. \cite{bassetti2024}. For the non-linear case, the derivation leverages a Gaussian equivalence \cite{mei2019, Gerace_2021, Aguirre-López2024random, Ariosto} informally justified via a class of generalized central limit theorems \cite{BM, NourdinQuantitative}.

\paragraph{Generalization error of FC networks with multiple outputs---}
Building from the effective theory reviewed in the previous section \cite{pacelli2023statistical}, we compute the posterior statistics of the outputs of the FC network defined in the problem setting. 
In particular, we are interested in the average generalization loss over an unseen example $x^0 \notin X$ with label $y^0$, that is $\epsilon_g(x^0, y^0) = \langle \Vert y^0 - f(x^0; \theta)\Vert^2\rangle$, where the notation $\langle \cdot \rangle$ indicates the expected value over the equilibrium Gibbs distribution $p_\beta(\theta \vert X, \mathbf y) = \mu(\theta) \mathrm{exp}\{-\beta\mathcal{L}(\theta)\}/Z$. 
As we explicitly check in the SM, the generalization loss can be reduced to the following expression: 
\begin{equation}
    \epsilon_\text{g}(x^0,y^0)=\Vert y^0-\Gamma\Vert^2+\Tr \Sigma \,, \label{eq:gen_err}
\end{equation}
where $\Gamma$ is the $D$-dimensional vector of average predictions $\Gamma_a = \langle f^a(x^0) \rangle$ , and $\Sigma$ is the $D \times D$ variance of the predictor $\Sigma_{ab} =\langle f^a(x^0) f^b(x^0) \rangle - \Gamma_a \Gamma_b$. In Fig. \ref{fig:1}, we show the outcome of numerical simulations to assess the validity of the previous formula in describing the behaviour of finite Bayesian networks. 
Considering a test set of $P_{\mathrm{test}}$ examples $X_{\mathrm{test}}=\{x^i\}$ with labels $Y_{\mathrm{test}} =\{y^i\}$, we compute the average empirical generalization loss after training, that is $ \epsilon^{\mathrm{emp}}_g = \sum_{i=1}^{P_{\mathrm{test}}}\Vert y^i-f(x^i)\Vert^2/P_{\mathrm{test}}$, for three different datasets (synthetic random, MNIST, CIFAR10) and compare it with theoretical prediction in Eq. \eqref{eq:gen_err}.
To enforce sampling from the desired Gibbs posterior, we train our models using the discretized Langevin equation. See the SM for full details on Bayesian experiments and the datasets used. 

Despite the overall very good agreement between experiments and predictions, it is worth stressing that deviations can be observed for small values of $N_1$, which is expectable since the predictions hold in the thermodynamic limit. In the SM, we perform preliminary analysis on the random projected MNIST dataset, to quantify the possible sources of mismatch. Note that the proportional theory correctly reduces to the infinite-width one for large values of $N_1$ compared to $P$. In Fig. \ref{fig:1}, we also show the predicted generalization loss obtained using a heuristic solution for the order parameter $Q$, obtained with a perturbative expansion around $Q = \mathbb 1 $. Unlike the exact solution $Q^*$, that we obtain by numerical minimization of the effective action, the perturbative heuristic solution is obtained analytically, as we explicitely show in the SM. Although the generalization loss obtained with the exact solution predicts the experimental one more accurately, the heuristic one provides a good approximation at a much lower computational cost. Additional numerical investigation is provided in the SM, Fig.~\ref{fig:supp:genErr}.

\begin{figure}[!ht]
\centering
    \includegraphics[width=.5\textwidth]
    {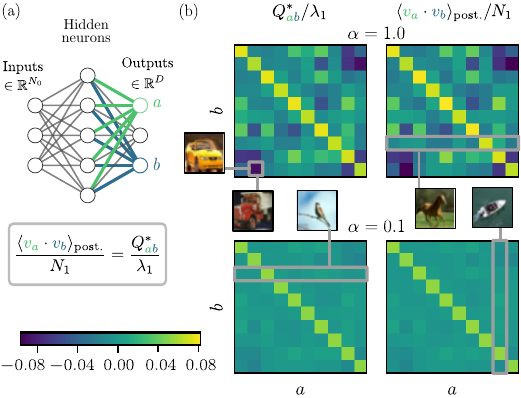}
    \caption{ \textbf{Kernel shape renormalization captures the effect of training finite Bayesian 1HL NNs with multiple outputs.} (a) Sketch of a FC network with $D = 3$ outputs, where we highlight weight vectors $v_a$ (green) and $v_b$ (blue). The predicted value for the quantity $v_a \cdot v_b /N_1$ after training is given by $Q^*_{ab}/\lambda_1$. (b) Box plots of the predicted matrix $Q^*/\lambda_1$ (left) vs box plot of the experimental value $vv^T$ of a network trained on the grayscaled CIFAR10 dataset ($N_0 = 28^2$ ; $D = 10$). We set $P=P_\textrm{test} = 1000$, $T = 0.01$, $\lambda_0 = \lambda_1 = 1.0$, and divide diagonal values by $20$.
    } 
\label{fig:2}
\end{figure}

\paragraph{Data-induced output-output correlations are predicted by the proportional-width theory---}The generalization loss is a measure of the network's predictive performance, but does not offer insight on its hidden representation of train examples. Here, we quantify data-induced output-output correlations by computing the $D \times D$ matrix observable $\langle v v^T \rangle$, whose entries measure the average overlap between weight vectors in the last layer that are associated with different output neurons. This computation follows through in the proportional regime, as we show in the SM, yielding: 
\begin{equation}
    \frac{\langle v v^T \rangle}{N_{1}}=\frac{Q^{*}}{\lambda_{1}}\, .
    \label{eq:overlap_result}
\end{equation}
The off-diagonal elements of $Q^*$ now possess a straightforward physical interpretation: they measure correlations between weight vectors $v_a$ and $v_b$, respectively the $a$-th and $b$-th row of matrix $v$ (see also the sketch in Fig. \ref{fig:2}).
Note that these correlations vanish in  the infinite-width limit, where $Q_{ab}^* = \delta_{ab}$, as was already noted in classic literature \cite{Neal, mackay1998introduction}.

In the case of one-hot encoding of labels, where the regression problem is closely related to multiclass classification, the role of such correlations can be understood with the following argument. If the network has learned meaningful hidden representations, two data points $x$ and $x'$ belonging to different classes $a$ and $b$ that share fundamental features (eg. “cars" and “trucks") 
will have similar internal representations in terms of pre-activations $h(x) \sim h(x')$. To correctly discriminate two such objects, the weight vector $v_a$ needs to be negatively correlated to weight vector $v_b$. 

\paragraph{Comparison between predicted and experimental correlations---}
We performed numerical validation of Eq. \eqref{eq:overlap_result}, whose results are displayed in Fig. \ref{fig:2}. We analyze the posterior experimental overlap between weight vectors $\langle v_a \cdot v_b \rangle$, and we compare it to the theoretical prediction in Eq. \eqref{eq:overlap_result} for two values of the load $\alpha = P/N_1 =  0.1, 1$, finding quantitative agreement in both cases. In the former case, closer to the infinite-width limit $\alpha \to 0$, we observe very small off-diagonal terms, as is expected in that regime. In the latter case, when $P = N_1$, non-trivial correlations emerge between different weight vectors. For instance, we observe that the classes “cars"/“1", “ships"/“8" and “trucks"/“9" of CIFAR10, all belonging to the macro-category of vehicles, display strongly negative overlaps of their correspondent weight vectors $Q^*_{19} \sim Q^*_{18} \sim Q^*_{98} = -0.1\,  N_1$ (see Fig. \ref{fig:2}, where we highlight $Q_{98})$. We report additional numerical simulations in the SM, Fig.~\ref{fig:supp:vv_as_orderP}, showing the behavior of predicted and experimental correlations as function of the number of hidden neurons $N_1$.

\begin{figure}[!ht]
\centering
\includegraphics[width=.45\textwidth]
    {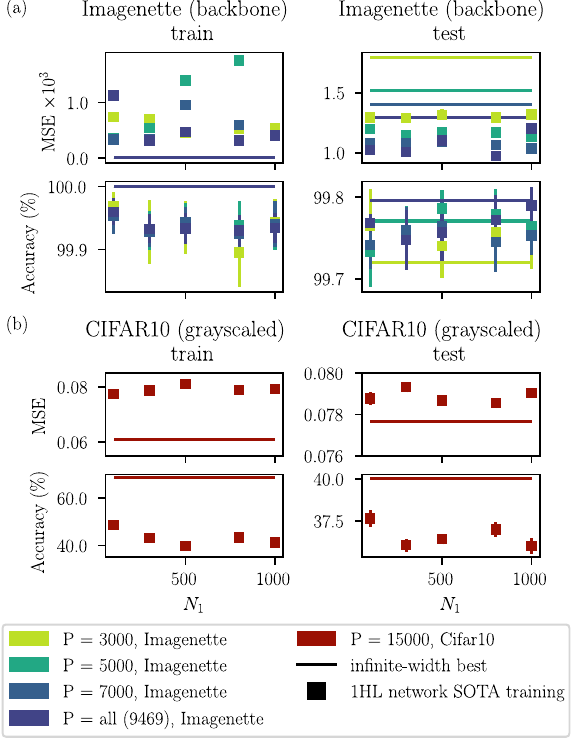}
    \caption{ \textbf{Bayesian training of FC networks is compatible with SOTA algorithms in terms of training and generalization performance.} MSE and accuracy of a 1HL network with ReLU activation for train (left) and test (right) sets as a function of the hidden layer size for different values of $P$. In all panels, dots represent the best performance of a 1HL networks trained with Adam, solid lines represent the best infinite-width predictions. The error bars indicate one standard deviation from the average over different network initializations. We show results for two computer-vision datasets: (a) Imagenette (see main text) preprocessed with a pre-trained backbone, and (b) gray-scaled CIFAR10. See SM for more details on the datasets and our pipeline to find the best model.
} 
\label{fig:3}
\end{figure}

\paragraph{Bayesian training of FC networks is comparable to SOTA optimization for realistic computer-vision datasets---} 
One major difference between real-world and Bayesian NNs is the choice of optimization algorithm. Bayesian training (Gibbs sampling) is not a popular choice among practitioners, mainly because of computational cost: state-of-the-art optimization algorithms like Adam are fast-converging, unlike Monte Carlo sampling, and yield solutions with good generalization properties \cite{adam}. 
To explore this possible mismatch, we performed numerical simulations with 1HL FC networks with ReLU activation trained with the Adam optimizer and MSE loss, where we carefully fine-tune hyperparameters (eg. $L^2$ regularizer, learning rate) to select the best possible model (see SM for more details). We compare this optimal performance with the infinite-width best performance, obtained with a grid hyperparameter search on $T, \lambda_0, \lambda_1$.
The results of such investigation are shown in Fig. \ref{fig:3} for two computer-vision datasets (Imagenette \cite{imagenette}, a subsample of ImageNet with $D= 10$, and CIFAR10). We find the expected compatibility between the two optimizers in terms of generalization performance, with Adam yielding slightly worse performances on CIFAR10 at $\alpha \sim 15$.
These observations are compatible with empirical studies made on infinite-width networks \cite{neurips_empirical_study}, where the authors show how FC networks are ineffective at finite-width, in the sense that they are eventually outperformed by an infinite-width model with a suitable choice of hyperparameters. A theoretical argument to understand this fact in single-output networks can be found in \reference{aiudi2023}. Note that the equivalence between Bayesian training and SOTA optimization breaks down when the input size $N_0$ is much smaller than the trainset size $P$, as we show in the SM for a synthetic dataset with $N_0 = 12$.

\paragraph{Discussion---}
In this work, we explored output-output correlations in finite Bayesian FC shallow networks, and their relation to the kernel shape renormalization mechanism in the proportional regime. In this setting, the existence of common features extracted by hidden units emerges as data-induced correlations between weighs of the last layer, which in turn induce dependencies among different output neurons. Remarkably, the matrix order parameter $Q^*$, that produces the NNGP kernel shape renormalization, quantifies such correlations in a straightforward way, measuring the overlap between weight vectors associated to different output neurons. This non-trivial behavior is in contrast to the simplistic lazy-training infinite-width, where feature learning is effectively suppressed and output-output correlations vanish together with off-diagonal elements of the matrix $Q^*$. We have conducted numerical experiments with finite models across synthetic and real-world datasets, finding quantitative match with our predictions, both in terms of generalization performance and weight correlations after training.  
It is worth stressing that our theory suffers some limitations, for instance, we are not yet able to control finite-width effects that generate mismatch between theory and experiments in the small $N_1$ regime. 
Interestingly, our experiments with FC ReLU networks show that the infinite-width prediction is competitive with Adam in terms of generalization, at least in the regime $P \sim N_1$. This fact has two main implications (i) it suggests that some of the phenomenology occurring in SOTA training is grasped by the Bayesian investigation (ii) it points to the somewhat disappointing consequence that the kind of feature learning that occurs in finite FC models is not very effective in improving generalization (at least in common computer vision datasets), as was already pointed out in recent literature \cite{neurips_empirical_study, aiudi2023}. As a final remark, we want to stress that there could be many other sources for the emergence of weight structure in real-world networks, not captured by our proportional-width investigation \cite{doi:10.1073/pnas.2311805121, fischer24critical, NEURIPS2021_b24d2101, seroussi2023natcomm, hanin2024bayesian}. For instance, in the heavily underparameterized limit $ P \sim N_0 N_1 $, FC networks can display highly non-trivial localization of the weight structure with the data-driven emergence of convolutional kernels \cite{ingrossoConv}, which cannot be taken into account by the effective theory presented in our manuscript. 

\emph{Acknowledgements}
R.B., R.P., P.R. and P.B. are supported by $\#$NEXTGENERATIONEU (NGEU). R.B. and P.R. are funded by the Ministry of University and Research (MUR), National Recovery and Resilience Plan (NRRP), project MNESYS (PE0000006) ``A Multiscale integrated approach to the study of the nervous system in health and disease” (DN. 1553 11.10.2022). R.P. and P.B. are funded by MUR project PRIN 2022HSKLK9 and P2022A889F. 

\bibliography{biblio_new}

\clearpage
\onecolumngrid
\setcounter{section}{0}
\renewcommand{\thesection}{\Roman{section}}
\renewcommand{\thesubsection}{\Alph{subsection}}

{\LARGE 
\begin{center}
    Supplemental Material for \\
    \textbf{“Kernel shape renormalization explains output-output correlations in finite Bayesian one-hidden layer networks"}
\end{center} }
\tableofcontents

\section{Computation of the generalization error for a FC network with multiple outputs}
\begin{figure}[!t]
\centering
    \includegraphics[width=1.\textwidth]
    {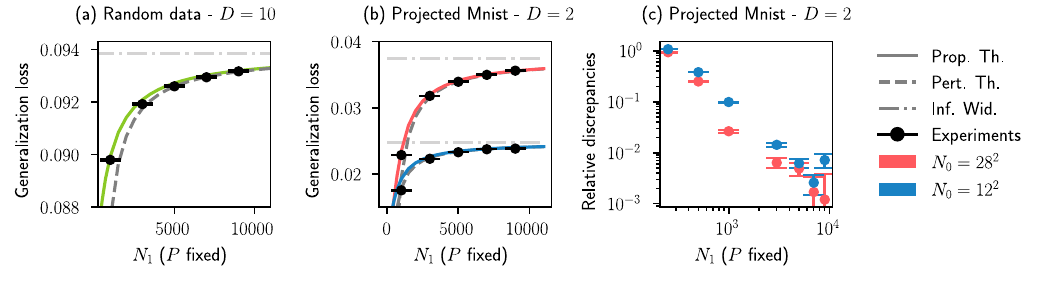}
    \caption{\textbf{Generalization loss of a fully-trained one-hidden-layer network with multiple outputs.}  
Panels (a) and (b) illustrate the generalization performance of a 1HL network with $D$ outputs as a function of the hidden layer size $N_1$ for two different datasets. Solid lines denote predictions from the effective Bayesian theory, dashed lines represent heuristic 1-loop perturbative solutions for the order parameters, dot-dashed lines indicate the infinite-width predictions, and points with error bars correspond to numerical experiments. (a) Results for random Gaussian data (see Sec.~\ref{sec:supp:details_datasets_be} for details), with $N_0 = 12^2$ input features and $D = 10$ output classes. (b) Results for the randomly projected MNIST dataset (see Sec.~\ref{sec:supp:details_datasets_be} for additional informations), with $N_0 = 28^2$ (red lines) and $N_0 = 12^2$ (blue lines) input features, and $D = 2$ output classes, where labels were hot-encoded. (c) Numerical investigation of the relative discrepancies between the effective theory and numerical predictions as a function of the hidden layer size $N_1$, for $N_0 = 12^2$ and $N_0 = 28^2$ using the projected MNIST dataset (see also Eq.~\ref{eq:supp:discrep} for details). Colors in panel (c) match those of panel (b). For all plots, we used $P = P_{\mathrm{test}} = 1000$, $\lambda_0 = \lambda_1 = 1.0$, $T = 0.01$, $\epsilon = 0.001$, and set biases to zero.} 
\label{fig:supp:genErr}
\end{figure}

For a new (unseen) example $(x^0, y^0) \notin X$, the generalization loss is defined as: 
\begin{equation}
    \Big \langle \Big\Vert y^0 - f_{\mathrm{FC}}(x^0)\Big\Vert^2 \Big \rangle_{p_\beta(\theta \vert X, \mathbf y)} = \sum_{a=1}^{D} \Big \langle \Big(y^0_a - f_{\mathrm{FC},a}(x^0)\Big)^2 \Big \rangle_{p_\beta(\theta \vert X, \mathbf y)} 
\end{equation}
where the index $a = 1 \ldots D $ runs over the network's outputs. We therefore need only compute the single-class generalization error. For the $k$-th class, this explicitly reads:
\begin{align}
    \Big \langle \Big(y^0_k - f_{\mathrm{FC},k}(x^0)\Big)^2 \Big \rangle_{p_\beta(\theta \vert X, \mathbf y)} & = \frac{1}{Z}\int \Bigg(\prod_{a, i_0, i_1=1}^{D, N_0, N_1} dw_{i_0 i_1}dv_{i_1 a} \mu(v_{i_1 a})\, \mu(w_{i_0 i_1})\Bigg)e^{- \frac{\beta}{2} \sum_{\mu a} (y^\mu_a - f_{\mathrm{FC},a}(x^\mu))^2 } \big(y^0_k - f_{\mathrm{FC},k}(x^0)\big)^2 \, ,
\end{align}
where the prior probability distributions $\mu(v_{i_1 a})$ and $\mu(w_{i_0 i_1})$ are Gaussian. The integral can be simplified by decoupling the contributions of different layers via Dirac delta functions:
\begin{align}
    \int \prod_{i_1 =1}^{N_1} \prod_{\mu = 0}^{ P} dh^\mu_{i_1} \delta\Big(h^\mu_{i_1}-\frac{1}{\sqrt{N_0}}\sum_{i_0}w_{i_1 i_0} x^\mu_{i_0}\Big) & = 1\, , \quad\quad\quad
    \int \prod_{a = 1}^{D} \prod_{\mu =0}^P ds^\mu_a \delta\Big(s^\mu_a-\frac{1}{\sqrt{N_1}}\sum_{i_1}v_{i_1 a} \sigma(h^\mu_{i_1}) \Big) = 1 \, ,
\end{align}
where we choose to run the index $\mu=0,1,\ldots, P$, to include all the trainset examples plus the test one $x^0$. When not explicitly stated, the index $\mu$ is understood to run from $1$ to $P$. By integrating over the weights, we obtain the following expression:
\begin{align}
    \label{eq:SM_gen_error_step1}
    \Big \langle \Big(y^0_k - f_{\mathrm{FC},k}(x^0)\Big)^2 \Big \rangle_{p_\beta(\theta \vert X, \mathbf y)} & = \frac{1}{Z} \int \Bigl( \prod_{a=1}^{D}\prod_{\mu=0}^{P} \frac{ds^\mu_a}{\sqrt{2\pi}} \frac{d\bar{s}^\mu_a}{\sqrt{2\pi}} \Bigr) e^{- \frac{\beta}{2} \sum_{\mu a} (y_a^\mu - s_a^\mu)^2 } e^{i\sum_{ a, \mu=0}\bar{s}^\mu_a s^\mu_a } \nonumber \\
    \times \Biggl[ \int \Big( \prod_{\nu=0}^{P} \frac{dh^\nu}{\sqrt{2\pi}} \frac{d\bar{h}^\nu}{\sqrt{2\pi}} \Big) &  ~ e^{i\sum_{\nu=0} \bar{h}^\nu h^\nu - \frac{1}{2}\sum_{\mu \nu = 0} \bar{h}^\mu \mathbb{C}^{\mu \nu} \bar{h}^\nu  - \frac{1}{2N_1\lambda_1} \sum_{a, \mu \nu=0} \bar{s}^\mu_a \sigma(h^\mu) \bar{s}^\nu_a \sigma(h^\nu) } \Biggr]^{N_1} (y^0_k - s^0_k)^2 \, .
\end{align}
In Eq.~\eqref{eq:SM_gen_error_step1}, the dependence on $i_1$ has been factorized, and the Fourier representation of the Dirac delta functions is used. In addition, we have introduced the $(P+1) \times (P+1)$ matrix $\mathbb{C}$, defined as $\mathbb{C}_{\mu\nu} = {x^\mu \cdot x^\nu}/{\lambda_0 N_0}$.
To perform the Gaussian integrals over $h^\mu$ and $h^0$, we insert a new set of variables $q_a$:
\begin{equation}
     q_a = \frac{1}{\sqrt{N_1 \lambda_1}} \sum_{\mu=0} \bar{s}^\mu_a \sigma(h^\mu) \, .
\end{equation}
This transformation leads to the problem of finding the following probability distribution:
\begin{align}
    \label{eq:SM_Pq}
    P(\{q\}) = \int \prod_{\nu=1}^{P} dh^\nu dh^0 \mathcal{N}_h(\boldsymbol{0}, \mathbb{C}) \prod_{a=1}^{D} \delta\Big( q_a - \frac{1}{\sqrt{N_1 \lambda_1}} \sum_{\mu=0} \bar{s}^\mu_a \sigma(h^\mu) \Big) \, .
\end{align}
In the proportional limit, we invoke a Gaussian equivalence for the probability distribution in Eq. \eqref{eq:SM_Pq}, heuristically justified via a set of generalized central limit theorems originally due to Breuer and Major \cite{BM, bardet2013, NourdinMulti}: 
\begin{equation}
    P(\{q\}) \rightarrow \mathcal{N}_q(\mathbf{0}, \Sigma(\bar s)) \quad\quad \text{if} \quad\quad P, N_1 \to \infty \, , \alpha=\text{fixed}\, ,
\end{equation}
with covariance matrix
\begin{align}
    \Sigma_{ab}(\bar s) = \frac{1}{N_1 \lambda_1} \int \prod_{\nu=1}^{P} dh^\nu dh^0   \frac{1}{\sqrt{(2\pi)^{P+1}\det \mathbb{C}}} ~ e^{ - \frac{1}{2}\sum_{\mu \nu = 0} h^\mu (\mathbb{C}^{-1})^{\mu \nu} h^\nu} \sum_{\mu \nu=0} \bar{s}^\mu_a \sigma(h^\mu) \bar{s}^\nu_b \sigma(h^\nu)  = \frac{1}{N_1 \lambda_1} \sum_{\mu\nu=0} \bar{s}^\mu_a K^{\mu \nu} \bar{s}^\nu_b \, .
\end{align}
In the previous equation we employed the short-hand notation $K^{\mu \nu} \equiv \lambda_1  K^{\text{NNGP}}(x^\mu, x^\nu)$ to indicate the Neural Network Gaussian Process kernel function \cite{g.2018gaussian,LeeGaussian}, evaluated on pair of train/test set inputs. The quantity in square brackets in Eq. \eqref{eq:SM_gen_error_step1} becomes: 
\begin{align}
    \left[ \int \prod_a d q_a e^{ -\frac{1}{2} \sum_a q_a^2} \mathcal{N}_q(\mathbf{0}, Q(\bar s)) \right]^{-\frac{N_1}{2}} = \left[ \det(\mathbb 1 + \Sigma(\bar s)\right]^{-\frac{N_1}{2}} 
\end{align}
Here, one needs to introduce a $D \times D$ matrix of order parameters $Q$ (one for each element of the matrix $\Sigma(\bar s)$). One can do so directly employing a general formula for the Fourier transform of a wishart distribution due to Ingham \cite{Ingham_1933}:
\begin{equation}
\det\left(\mathbb{1}_{D}+\Sigma(\bar{s})\right)^{-\frac{N_{1}}{2}} \propto\int_{Q>0} \prod_{ab} \diff Q_{ab}(\det Q)^{\frac{N_{1}-D-2}{2}}e^{-\frac{1}{2}\Tr\Sigma(\bar s) Q-\frac{1}{2}\Tr Q} \, ,
\label{eq:ingh_sieg}
\end{equation}
where the integration is over an ensemble of positive semi-definite matrices. Using the relation in Eq. \eqref{eq:ingh_sieg}, the integrals over the $s^\mu$ and $\bar{s}^\mu$ become Gaussian and can be easily computed. At the leading order, we thus get
\begin{align}
    \label{eq:SM_gen_error_step2}
    \Big \langle \Big(y^0_k - f_{\mathrm{FC},k}(x^0)\Big)^2 \Big \rangle = \frac{1}{Z} \int \frac{dQ }{2\pi} ~ & e^{-\frac{N_1}{2}S_{\text{FC}}(Q )} \int \Bigg( \prod_{a=1}^{D} \frac{ds^0_a}{\sqrt{2\pi}} \frac{d\bar{s}^0_a}{\sqrt{2\pi}} \Bigg) e^{i\sum_a \bar{s}^0_a \Big( s^0_a - \sum_{\mu\nu cb}y^\mu_c \big( \big[\frac{\mathbb{1}_D \otimes \mathbb{1}_P}{\beta}  + K^{\text{(R)}} \big]^{-1}\big)_{cb}^{\mu\nu} (K^{\text{(R)}})_{ba}^{\nu 0}\Big) } \notag  \\
    \times ~ (y^0_k - s^0_k)^2  &  e^{-\frac{1}{2} \sum_{ac} \bar{s}^0_a \big[ (K^{\text{(R)}})_{ac}^{00} - \sum_{\mu\nu kt} (K^{\text{(R)}})_{ka}^{\mu 0}  \big( \big[\frac{\mathbb{1}_D \otimes \mathbb{1}_P}{\beta} + K^{\text{(R)}} \big]^{-1}\big)_{kt}^{\mu\nu} (K^{\text{(R)}})_{tc}^{\mu 0} \bigr] \bar{s}^0_c}\, .
\end{align}
Note that in Eq.~\eqref{eq:SM_gen_error_step2} we used the Renormalized Kernel notation
\begin{align}
    (K^{\text{(R)}})_{ab}^{\mu\nu} = \frac{Q_{ab}K^{\mu\nu}}{\lambda_1}\, , \quad\quad 
    (K^{\text{(R)}})_{ab}^{\mu0} = \frac{1}{\lambda_1}Q_{ab}K^{\mu0} \, ,\quad\quad (K^{\text{(R)}})_{ab}^{00} = \frac{1}{\lambda_1}Q_{ab}K^{00} \, .
\end{align}
Performing the remaining Gaussian integrations in Eq. \eqref{eq:SM_gen_error_step2}, and the saddle point approximation for the integration in $Q$, we are able to isolate the posterior distribution $p\left( f_{\mathrm{FC}}(x^0)\right) \sim \mathcal N(\Gamma, \Sigma)$, where the mean and covariance have the following components:
\begin{align}
    \label{eq:SM_bias}
    \Gamma_a & = \sum_{\mu\nu cb}y^\mu_c \Big( \Big[\frac{1}{\beta} \mathbb{1}_D \otimes \mathbb{1}_P + K^{\text{(R)}}_{Q^*} \Big]^{-1}\Big)_{cb}^{\mu\nu} (K^{\text{(R)}}_{Q^*})_{ba}^{\nu 0} \, ,\\
    \label{eq:SM_var}
    \Sigma_{ac} & =  (K^{\text{(R)}}_{Q^*} )_{ac}^{00} - \sum_{\mu\nu kt} (K^{\text{(R)}}_{Q^*} )_{ka}^{\mu 0}  \Big( \Big[\frac{1}{\beta} \mathbb{1}_D \otimes \mathbb{1}_P + K^{\text{(R)}}_{Q^*}  \Big]^{-1}\Big)_{kt}^{\mu\nu} (K^{\text{(R)}}_{Q^*} )_{tc}^{\nu 0}\, .
\end{align}
Here we emphasize that the Renormalized Kernel in Eqs.~\eqref{eq:SM_bias} and \eqref{eq:SM_var}, which depends on the order parameter $Q$, is evaluated at the saddle point $Q^*$. Finally, we recover the standard bias-variance decomposition of the generalization loss, given in Eq.~\eqref{eq:gen_err} of the main text:
\begin{align}
    \label{eq:SM_gen_error}
    \Big \langle \Big(y^0_k - f_{\mathrm{FC},k}(x^0)\Big)^2 \Big \rangle_{p_\beta(\theta \vert X, \mathbf y)} = \epsilon_k(x^0, y^0) = (y^0_k - \Gamma_k)^2 + \Sigma_{kk} \, .
\end{align}
In Fig.~\ref{fig:supp:genErr}, panels (a) and (b), we present additional numerical studies validating Eq.~\ref{eq:SM_gen_error} for finite values of $N_1$ and $P$. The numerical results demonstrate good agreement with the theoretical predictions (see also Fig.~\ref{fig:1} in the main manuscript). Furthermore, in panel (c), we show preliminary numerical investigations into potential sources of discrepancies between the numerical experiments and theoretical predictions. Defining the generalization error computed via Monte Carlo methods as $\epsilon_{g,\text{MC}}$ (see Sec.~\ref{sec:supp:MC}) and the corresponding error from the effective theory as $\epsilon_{g,\text{TH}}$, we calculate the relative discrepancy as 
\begin{equation}
    \label{eq:supp:discrep}
    \Delta = \Bigl| \frac{\epsilon_{g,\text{MC}} - \epsilon_{g,\text{TH}}}{\epsilon_{g,\text{TH}}} \Bigr|,
\end{equation}
which quantifies the relative deviation between theory and numerical experiments. As expected, the relative discrepancies decrease monotonically with an increasing number of hidden neurons $N_1$. Moreover, for a fixed value of $N_1$, the discrepancies decrease significantly as the number of input features $N_0$ increases. This observation suggests that additional systematic deviations also arise from the finite number of input features considered.

\section{Computation of the average overlap between weights of different output neurons \label{supp:sec:Qab}}

Denoting $a = 1 \ldots D$ the index that runs over the network's outputs, we are interested in the overlap between column $a$ and $b$ of matrix $v$, that we denote $v_a = \lbrace v_{ia} \rbrace_{i=1}^{N_1} $ and $v_b= \lbrace v_{ib} \rbrace_{i=1}^{N_1}$:
\begin{align}
    & \langle v_a \cdot v_b \rangle_{p_\beta(\theta \vert X, Y)} = \left. \frac{\partial \log \tilde Z}{\partial \lambda_{ab}} \right\vert_{\lambda_{ab} = \lambda_{ab}^*}\, , \label{eq:supp:vv}\\
    & \tilde Z=\int \diff w\,\diff v\, \mathrm{exp}\left( -\frac{\beta}{2} \tilde{\mathcal L}(w,v) \right)\, , \label{eq:ztilde}\\
    &\tilde{\mathcal L}(w,v) = \sum_{i} v_i^T  \Lambda v_i - N_1 \log \det \Lambda + \lambda_0 \Vert w\Vert^2 + \nonumber \\
    & \quad - N_0 N_1 \log \lambda_0 +\beta \mathcal L (w,v) \, \label{eq:Ltilde},
\end{align}
where the index $i = 1 \ldots N_1$ runs over the neurons in the hidden layer, $\Lambda$ is a $D \times D$ matrix with elements $(\Lambda)_{ab} = \lambda_{ab}$, and the result has to be evaluated on the usual ensemble given by $\Lambda^{*}= \lambda_{1}\mathbb{1}$. 

\begin{figure}[!t]
\centering
    \includegraphics[width=.99\textwidth]
    {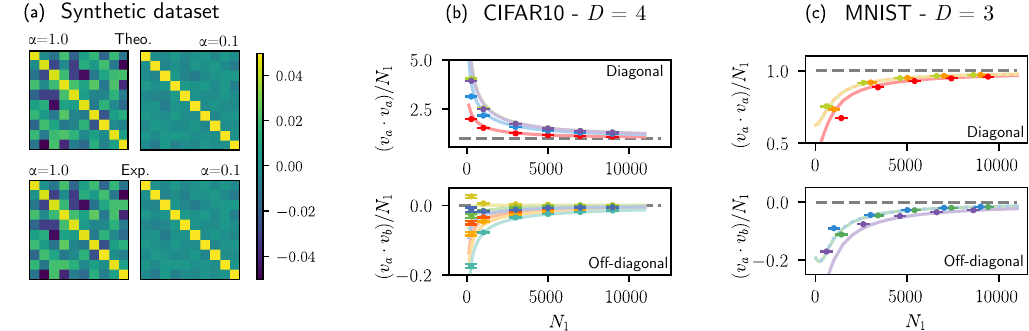}
    \caption{\textbf{Overlaps of the readout layer in a fully-trained one-hidden-layer network with multiple outputs.} Panel (a) displays the overlaps of the readout layer in a 1HL FC network with $D = 10$ classes for the synthetic Gaussian dataset, with $N_0 = 12^2$ input features. The first column corresponds to $\alpha = 1.0$, while the second column corresponds to $\alpha = 0.1$. The first row shows predictions from the Bayesian effective theory, while the second row reports numerical results for finite networks. The diagonal elements have been divided by $20$. Panels (b) and (c) present the overlaps as a function of the number of hidden neurons $N_1$ for two datasets: the CIFAR10 dataset with $D = 4$ classes (“cars," “birds," “cats," “deer") and the MNIST dataset with $D = 3$ classes (“$1$," “$2$," “$3$"). Solid lines represent predictions from the Bayesian effective theory, dashed lines denote the infinite-width predictions, and points with error bars correspond to numerical experiments. The color scheme ensures that predictions and numerical results for the same observable are consistently matched. Specifically, the diagonal elements of the covariance matrix are shown in the first row, while the off-diagonal elements are displayed in the second row (including their corresponding infinite-width limits). In panel (c), data points for the same $N_1$ are slightly offset for better visibility, with the central point indicating the exact value. For both the CIFAR10 and MNIST datasets, labels were hot-encoded. In all simulations, we used $P = P_{\mathrm{test}} = 1000$, $\lambda_0 = \lambda_1 = 1.0$, $T = 0.01$, $\epsilon = 0.001$, and set biases to zero.} 
\label{fig:supp:vv_as_orderP}
\end{figure}

With a computation similar to the ones in \cite{pacelli2023statistical}, the modified partition function in Eq. \eqref{eq:ztilde} can be reduced to a saddle point integral $\tilde Z= \int\diff Q\,\mathrm{exp}(-N_{1}\tilde S(Q) /2) $:
\begin{align}
\tilde S(Q)= & \Tr\,Q\Lambda-\Tr\log\,Q\Lambda + \nonumber  \\
& +\frac{\alpha_{1}}{P}\Tr\log\beta\left[\frac{1}{\beta}\;\mathbb{1}_{D}\otimes\mathbb{1}_{P}+K_{Q}^{\mathrm{(R)}}(X,X)\right]+ \nonumber\\
& + \frac{\alpha_{1}}{P}y^{\top}\left[\frac{1}{\beta}\;\mathbb{1}_{D}\otimes\mathbb{1}_{P}+K_{Q}^{\mathrm{(R)}}(X,X)\right]^{-1}y\, .
\end{align}
The renormalized kernel reads: 
\begin{equation}
    K_{Q}^{\mathrm{(R)}}(X,X) = Q \otimes \tilde{K}^{\mathrm{NNGP}} (X,X) \, ,
    \label{eq:supp:kernel2}
\end{equation}
where $Q$ is a $D \times D$ matrix, the symbol $\otimes$ denotes the tensor product, and $\lambda_1$ is the prior variance of the readout (last layer) weights. Note that, in Eq. \ref{eq:supp:kernel2}, the NNGP kernel $\tilde K$ is defined as the NNGP one, short of the Gaussian prior \cite{LeeGaussian}. It is easy to show that: 
\begin{equation}
\frac{\partial\log \tilde Z}{\partial\lambda_{ab}}=  \frac{N_{1}}{2} \left.\Tr \left[ \Lambda^{-1} \Delta_{ab} \right] \right\vert_{\Lambda = \Lambda^*} - \frac{1}{2}\langle v_{a}\cdot v_{b}\rangle \, ,
\end{equation}
where $\Delta_{ab}$ are $D\times D$ matrices with $(\Delta_{ab})_{ij} = \delta_{ai}\delta_{bj}$. When the partition function is expressed in terms of $\tilde S$ we have: 
\begin{align}
    &\frac{\partial\log \tilde Z}{\partial\lambda_{ab}}= -\frac{N_1}{2} \left.\frac{\partial \tilde S}{\partial\lambda_{ab}}\right\vert_{Q = Q^{*},\Lambda = \Lambda^{*}}\, , \\
    & \frac{\partial \tilde S}{\partial\lambda_{ab}}=\left.\Tr\left[Q\Lambda^{-1}\Delta_{ab}\right] \right\vert_{Q =Q^*, \Lambda = \Lambda^*}- \left. \Tr\left[\Lambda^{-1}\Delta_{ab}\right] \right\vert_{\Lambda = \Lambda^*}\, .
\end{align}
Combining the previous expression, we obtain:
\begin{equation}
    \label{eq:supp:vv_as_orderP}
    \frac{\langle v_{a}\cdot v_{b}\rangle}{N_{1}}=\frac{Q_{ab}^{*}}{\lambda_{1}} \, .
\end{equation}
This result has a natural interpretation in terms of Gaussian Processes, as was shown in \cite{pacelli2024GP}. We provide additional numerical validation of Eq.~\eqref{eq:supp:vv_as_orderP} for finite Bayesian networks in Fig.~\ref{fig:supp:vv_as_orderP}. In panel (a), we display the matrix of readout overlaps for two values of $\alpha = 0.1$ and $1.0$ for the synthetic Gaussian dataset. Panels (b) and (c) present numerical results for the readout overlaps against the order parameter at the saddle point, plotted as function of the number of hidden neurons $N_1$, for both the CIFAR10 and MNIST datasets. In all cases, we observe good agreement between the numerical Bayesian simulations and the predictions of the effective theory.

\section{Details of numerical experiments with Langevin dynamics}
\label{sec:supp:MC}

To perform Bayesian training, we employed the Langevin Monte Carlo (LMC) algorithm. LMC is effective in this context because configurations generated by the Langevin equation are distributed according to the stationary solution of the Fokker-Planck equation, ensuring correct sampling from the Bayesian posterior. The Langevin equation is given by:
\begin{equation}
    \dot{\theta}_i(t) = - \Biggl( \frac{\partial \tilde{\mathcal L}}{\partial \theta_i}\Biggr)_{\theta(t)}  + \sqrt{2T} \eta_i(t),
\end{equation}
where $\tilde{\mathcal L} $ is the effective loss that takes Gaussian priors and temperature into account: 
\begin{equation}
    \tilde{\mathcal{L}} = \mathcal L + \frac{T\lambda_0}{2} \Vert w \Vert^2 + \frac{T\lambda_1}{2} \Vert v \Vert^2 \, ,
\end{equation}
and $\eta_i(t)$ represents a Gaussian noise term with zero mean and unitary diagonal correlations
\begin{equation}
    \langle \eta_i(t_1) \rangle = 0, \quad \text{and} \quad \langle \eta_i(t_1) \eta_j(t_2) \rangle = \delta_{ij} \delta(t_1 - t_2).
\end{equation}
In practice, this equation is numerically integrated using a finite time step $\epsilon$, leading to the discrete update rule:
\begin{equation}
    \label{supp:eq:langevin}
    \theta_i(t + \epsilon) = \theta_i(t) - \epsilon \Biggl( \frac{\partial \tilde{\mathcal L}}{\partial \theta_i}\Biggr)_{\theta(t)} + \sqrt{2\epsilon T} \eta_i(t).
\end{equation}
Although discretization introduces systematic errors—since the exact Boltzmann distribution is not perfectly sampled—the errors scale as $\mathcal{O}(\epsilon^n)$, where $n$ is the order of the integration scheme. For our simulations, we used the first-order Euler integrator ($n = 1$), which provides a simple yet effective method, especially compared to the more conventional Metropolis algorithm. The trade-off is that LMC requires calculating the gradient of the loss function at each step, but it benefits from proposing non-local updates \cite{Barbu2020}. When using the Euler integrator, finite learning rate effects manifest as first-order corrections, requiring linear extrapolations as $\epsilon \to 0$, which may demand additional simulations. In this study, an alternative approach is employed: simulating the system with sufficiently small step sizes to ensure that systematic effects are negligible compared to statistical uncertainties. This can be validated by further reducing the learning rate and comparing the outcomes. While this method involves working with very small step sizes, which could lead to significant autocorrelations, the large number of available configurations has ensured that these autocorrelations are effectively controlled.

By iterating the update rule in Eq.~\eqref{supp:eq:langevin}, we generate a series of neural network weight configurations. For the experiments reported here, we ran the LMC simulations for roughly $\approx 50 \cdot 10^6$ steps, saving the generalization error and the covariances of the readout layer only once every 100 updates. This procedure reduces computational costs associated with input/output latency without losing statistical precision, provided the saving step is shorter than the autocorrelation time.

The initial weights were drawn from a normal distribution with zero mean and unit variance. Since these initial configurations are far from the equilibrium configurations, the early Monte Carlo updates were discarded during a thermalization phase. We visually inspected the simulation results and determined a default thermalization cutoff of 5000 steps (though this value was adjusted when necessary). Data collected before the system reached equilibrium were excluded from further analysis. 

\begin{figure}[!t]
\centering
    \includegraphics[width=.98\textwidth]
    {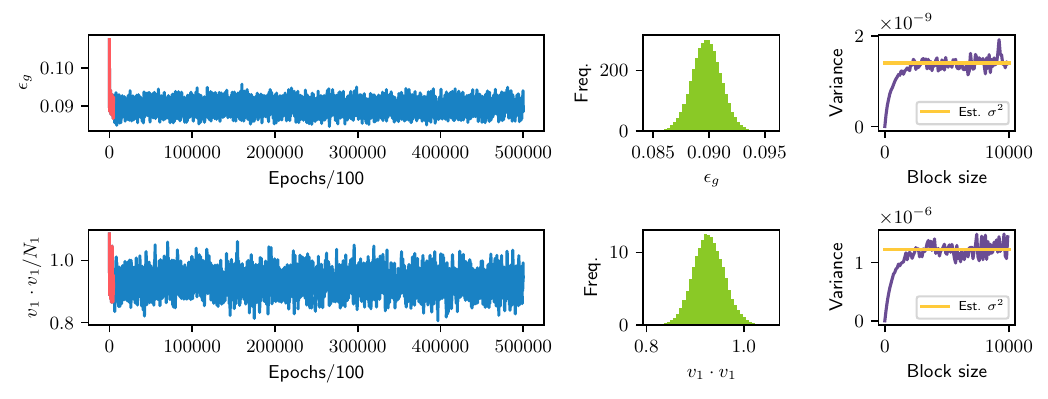}
    \caption{\textbf{Analysis of Monte Carlo Data for Generalization Loss and Readout Overlaps.} We illustrate Monte Carlo results for the generalization loss $\epsilon_g$ (first row) and the normalized overlap $v_1 \cdot v_1 / N_1$ (second row) in a 1HL FC network with $D = 10$ output classes. The dataset consists of random Gaussian inputs, with labels generated using a non-linear 1HL neural network as the teacher (see Sec.~\ref{sec:supp:details_datasets_be} for details), and $N_0 = 12^2$ input features. Plots in the first column illustrate the transient thermalization phase (red solid lines) at the start of the simulation and the thermalized Monte Carlo signal (blue solid lines). The signal is displayed in units of the learning rate $\epsilon = 0.001$. The central column shows histograms of the empirical Monte Carlo distributions of the two observables, with areas normalized to unity. Note that Monte Carlo distributions are not generally expected to be Gaussian (e.g., for the generalization error of a single unseen example). However, the tendency toward Gaussianity observed here arises from averaging over multiple unseen examples, as we set $P_{\text{test}} = 1000$, and multiple weights. The last column reports the determination of the statistical error on the mean. Block sizes are expressed in terms of the saving step. The variance of the mean, shown as yellow horizontal lines, is calculated as the average of the blocked variances (violet solid lines) in the plateau region. In all plots, we use $P = 1000$, $\lambda_0 = \lambda_1 = 1.0$, $T = 0.01$, and set the biases to zero.} 
\label{fig:supp:MC}
\end{figure}

As is standard practice in Monte Carlo simulations, we computed the generalization error in Eq.~\eqref{eq:gen_err} and the readout layer covariance matrix in Eq.~\eqref{eq:overlap_result} (see the main manuscript) using thermalized network configurations from the time-averaged results \cite{Binder2002}, namely:
\begin{align}
    \epsilon_g(x^0, y^0) & \approx \frac{1}{T} \sum_{t=1}^{T} \epsilon_g (\theta(t); x^0, y^0), \\
    \langle v_a \cdot v_b \rangle & \approx \frac{1}{T} \sum_{t=1}^{T} v(t)_{a} \cdot v(t)_{b},
\end{align}
where $T$ is the number of total configurations $\theta(t)$, $t=1,\ldots,T$, generated in the simulation. Each $\epsilon_g(\theta(t); x^0, y^0)$ and $v(t)_{a} \cdot v(t)_{b}$ corresponds to the squared error and readout layer covariances evaluated for a particular network configuration $\theta(t)$. In Fig.~\ref{fig:supp:MC} (first and second columns), we present an example of Monte Carlo data for both the generalization error and the readout overlaps. As long as the stochastic process generating these configurations reaches the canonical distribution $P[\theta] = e^{-\beta \tilde{\mathcal{L}}(\theta)} / Z$, the time-averaged estimate converges to the true expectation value. Under general conditions, the discrepancy between the time average and the true value scales as $1/\sqrt{T}$.

To estimate the statistical errors, we employed the blocking method, a standard technique for analyzing correlated data. This method involves dividing the data into blocks and averaging within each block, which reduces the impact of correlations between successive samples. As the block size increases, the error estimate becomes more accurate, eventually stabilizing when the blocks are large enough to be uncorrelated (see, for example, the last column of Fig.~\ref{fig:supp:MC}). In addition to the blocking method, we verified our results using the Gamma Function method \cite{Wolff2004}, obtaining consistent estimates of the statistical errors in both cases.

\subsection{Details on datasets for Bayesian experiments}
\label{sec:supp:details_datasets_be}
The datasets used in this study for Bayesian experiments include MNIST, CIFAR10, Randomly Projected MNIST, and Synthetic datasets. Denoting the number of input features as $N_0$ and the number of output classes as $D$, the datasets were constructed as follows:
\begin{itemize}
    \item \textbf{CIFAR10 and MNIST datasets} \\
    We used standard versions of the (grayscale) CIFAR10 and MNIST datasets, where all input features were flattened and normalized to have zero mean and unit variance. To restrict classification tasks to a subset of $D$ classes, a filtering procedure was applied to retain only samples corresponding to the selected classes. For reproducibility, a deterministic random seed was used to ensure reproducibility when selecting samples, $P$ for training and $P_{\text{test}}$ for testing. Finally, the categorical labels for the $D$ classes were converted into one-hot encoded representations.
    \item \textbf{Synthetic datasets} \\
    Synthetic input data for both the training and test sets are generated as random samples drawn from a normal distribution. Specifically, the training dataset $X = \{ x^\mu \in \mathbb{R}^{N_0} \}_{\mu=1}^P$ and the test dataset $X_{\text{test}} = \{ x^i \in \mathbb{R}^{N_0} \}_{i=1}^{P_{\text{test}}}$ are created by drawing each sample from a multivariate normal distribution with zero mean and unit variance:
    \begin{align}
        x^\mu & \sim \mathcal{N}(0, I_{N_0}) \quad \text{for} \quad \mu = 1, \dots, P, \\
        x^i & \sim \mathcal{N}(0, I_{N_0}) \quad \text{for} \quad i = 1, \dots, P_{\text{test}}.
    \end{align}
    The model used to generate the train output labels $Y = \{ y^\mu \in \mathbb{R}^{D} \}_{\mu=1}^P$ and the test output labels $Y_{\text{test}} = \{ y^i \in \mathbb{R}^{D} \}_{i=1}^{P_{\text{test}}}$ is a simple feedforward neural network with one hidden layer and multiple outputs, namely
    \begin{align}
        y^\mu_a = \frac{1}{\sqrt{N_1}} \sum_{i_1=1}^{N_1} v_{i_1 a} \sigma\left( \frac{1}{\sqrt{N_0}} \sum_{i_0=1}^{N_0} w_{i_1 i_0} x^\mu_{i_0} \right), \quad \text{for} \quad a=1,\ldots,D \quad \mu=1,\ldots, P \, , \\
        y^i_a = \frac{1}{\sqrt{N_1}} \sum_{i_1=1}^{N_1} v_{i_1 a} \sigma\left( \frac{1}{\sqrt{N_0}} \sum_{i_0=1}^{N_0} w_{i_1 i_0} x^i_{i_0} \right), \quad \text{for} \quad a=1,\ldots,D \quad i=1,\ldots, P_{\text{test}}\, .
    \end{align}
    The weights of both the first and second layers were initialized using a standard normal distribution with zero mean and unit variance. For all simulations, the activation function was chosen as \( \text{Erf} \) (error function), and the number of hidden neurons was set to \( N_1 = 1000 \).
    \item \textbf{Randomly projected MNIST} \\
    In this case, the dataset was generated by applying a random projection to the original MNIST dataset. The raw input images, initially represented as vectors in a $\tilde{N}_0$-dimensional feature space, were transformed using a random projection matrix $\Pi$ of dimensions $N_0 \times \tilde{N}_0$. Each element  $\Pi_{ij}$ was independently drawn from a normal distribution with zero mean and unit variance. The transformation is expressed as:  
    \begin{equation}
    x_{\text{proj}} = \text{ReLU}\left( \frac{1}{\sqrt{\tilde{N}_0}} \Pi x \right),
    \end{equation}
    where \( N_0 \) represents the number of input features in the projected space. This process was applied separately to both the train and test datasets. While the transformed inputs reside in the lower-dimensional space defined by \( N_0 \), the original (one-hot encoded) labels from the MNIST dataset were preserved.  
\end{itemize}

\section{Heuristic solutions to the Saddle Point equations: 1-loop approximation of the order parameters}

The computation of observables in the effective theory, such as predictions, generalization loss, etc., requires determining the value of the order parameters at the saddle-point solution. To locate the minimum around which the saddle point concentrates, we need to numerically minimize the effective action. This procedure requires evaluations of the gradient of the action, which, in turn, necessitates multiple inversions of (often large)  $P \times P$  matrices. This can be accomplished in different cases (see, for example, Ref.~\cite{baglioni2024predictive} or Fig.~\ref{fig:1} of the main text, solid lines). Here, we present a simple method to approximate the values of the order parameters by implementing perturbation theory in the small parameter  $\alpha_1$, considering small deviations around the infinite-width limit. Roughly speaking the procedure involves solving the non-perturbative saddle-point equations iteratively using perturbation theory, \textit{i.e.} considering expansions of the order parameters in the form:
\begin{equation}
    \label{eq:expansionQ}
    Q = Q^{(0)} + \alpha_1 Q^{(1)} + \alpha_1^2 Q^{(2)} + \ldots \, .
\end{equation}
The full, non-perturbative saddle-point equation is expressed as
\begin{align}
    \label{eq:supp:full_deriv}
    \frac{\partial S_{\text{FC}}}{\partial Q_{ij}} = \delta_{ij} - & \Tr (Q^{-1} \mathbb{E}_{ij}) + \frac{\alpha_1}{P} \Tr \Bigl\{ \Bigl[ \frac{1}{\beta} \mathbb{1}_D \otimes \mathbb{1}_P + \frac{1}{\lambda_1} Q \otimes K \Bigr]^{-1} \frac{\mathbb{E}_{ij} \otimes K}{\lambda_1} \Bigr\} \\
    & -\frac{\alpha_1}{P}\mathbf y^T \Bigl[ \frac{1}{\beta} \mathbb{1}_D \otimes \mathbb{1}_P + \frac{1}{\lambda_1} Q \otimes K \Bigr]^{-1} \Bigl( \frac{\mathbb{E}_{ij} \otimes K}{\lambda_1} \Bigr) \Bigl[ \frac{1}{\beta} \mathbb{1}_D \otimes \mathbb{1}_P + \frac{1}{\lambda_1} Q \otimes K \Bigr]^{-1}\mathbf y\nonumber
\end{align}
where $S_{\text{FC}}$ is defined in Eq.~\eqref{eq:S_FC} of the main text and
\begin{equation}
    \label{eq:supp:E}
(\mathbb{E}_{ij})_{kl} = \begin{cases}
    1, & \text{if } (k,l) = (i,j), \\
    1, & \text{if } (k,l) = (j,i), \\
    0, & \text{otherwise}.
\end{cases}
\end{equation}
It is worth noting that the introduction of the symmetric matrices in Eq.~\eqref{eq:supp:E} guarantees the symmetric nature of the derivative of the effective action. 
Considering only $1$-loop corrections, we observe that in Eq.~\eqref{eq:supp:full_deriv}, the inverse of the matrix Q can be expanded in perturbation theory as follows:
\begin{align}
    \Bigl( Q^{(0)} + \alpha_1 Q^{(1)} + \ldots \Bigr)^{-1} & = ( Q^{(0)} )^{-1} - \alpha_1 ( Q^{(0)} )^{-1} Q^{(1)} ( Q^{(0)} )^{-1} + \ldots \, .
\end{align}
Using this expansion, we land with the zero-order equation:
\begin{equation}
    \delta_{ij} - \Tr \Bigl( ( Q^{(0)} )^{-1}\mathbb{E}_{ij} \Bigr) = 0\, ,
\end{equation}
the solution of which gives the infinite-width result $Q^{(0)} = \mathbb{1}$. The $1$-loop equation is obtained by considering first-order corrections in $\alpha_1$ along with the zero-order solution. The first-order equation is thus given by:
\begin{align}
    \label{eq:supp:final_perturbative_eq}
    \Tr \Bigl( Q^{(1)} \mathbb{E}_{ij} \Bigr) = - \frac{1}{P} \Tr \Bigl\{ \Bigl[ \frac{1}{\beta} \mathbb{1}_D \otimes \mathbb{1}_P + &  \frac{1}{\lambda_1} \mathbb{1}_D \otimes K \Bigr]^{-1} \frac{\mathbb{E}_{ij} \otimes K}{\lambda_1} \Bigr\} \\
    & +\frac{1}{P}\mathbf y^T \Bigl[ \frac{1}{\beta} \mathbb{1}_D \otimes \mathbb{1}_P + \frac{1}{\lambda_1} \mathbb{1}_D \otimes K \Bigr]^{-1} \Bigl( \frac{\mathbb{E}_{ij} \otimes K}{\lambda_1} \Bigr) \Bigl[ \frac{1}{\beta} \mathbb{1}_D \otimes \mathbb{1}_P + \frac{1}{\lambda_1} \mathbb{1}_D \otimes K \Bigr]^{-1}\mathbf y\nonumber \, .
\end{align}
Note that Eq.~\eqref{eq:supp:final_perturbative_eq} provides an explicit equation for $Q^{(1)}$, as the trace and matrix multiplication with $\mathbb{E}_{ij}$ select a specific element $Q^{(1)}_{ij}$. Unlike the numerical minimization of the full effective action, determining the perturbative order parameters involves only matrix products, with no additional iterative computations. 

\section{Numerical experiments with Adam optimizer}
The numerical experiments involving training 1HL fully connected NNs with the Adam optimizer were conducted using a combination of data preprocessing and hyperparameter optimization. For the Imagenette dataset, a subsample of 10 ImageNet classes (tench, English springer, cassette player, chain saw, church, French horn, garbage truck, gas pump, golf ball, parachute), which possesses high inherent complexity, a pretrained EfficientNet backbone \cite{efficient_net_ICML} was used prior to the 1HL structure.

A backbone, in the context of deep learning, refers to a pre-trained neural network, often trained on large datasets, which can then be adapted to different tasks. This pre-trained network serves as a feature extractor, capturing general patterns and structures in the data, such as certain edge or color related features, that are common in visual datasets.

Using a backbone is particularly useful in cases where datasets are complex or limited in size, as it allows the model to leverage knowledge from extensive, pre-existing training. This kind of knowledge transfer, as opposed to training networks from scratch, has indeed become standard among practitioners. The backbone we used was pre-trained on the full ImageNet dataset.  

For the other computer vision dataset that we employed, CIFAR-10 \cite{krizhevsky2009learning}, the preprocessing has been limited to gray-scaling. (CIFAR-10, California Housing, and Multiple Output Regression), no backbone was employed. Additionally, all data were normalized to achieve zero mean and unit variance.  

\subsection{Hyperparameter Optimization}
Each model and dataset underwent a hyperparameter optimization procedure to optimize the learning rate, $L_2$ regularization strength, and batch size. This was done using Ray Tune \cite{liaw2018tune}, a tool for distributed hyperparameter optimization. 

The hyperparameter search leveraged Ray Tune’s framework with an \texttt{AsyncHyperBandScheduler}, which efficiently allocates resources by prioritizing promising hyperparameter configurations. Using \texttt{HyperOptSearch}, we explored learning rate, $L_2$ regularization strength, and batch size to minimize the validation mean squared error, and the best combination was then selected. The corresponding model underwent a full earlystopped training. All estimations of out-of-sample error were based on a validation set comprising around 20\% of the training data.

Following the hyperparameter optimization process, 20 independent training realizations were conducted for each one of the presented model, and the mean and variance of the test loss were calculated and reported.

\subsection{Low-dimensional data}
\begin{figure}[!ht]
\centering
\includegraphics[width=.55\textwidth]
    {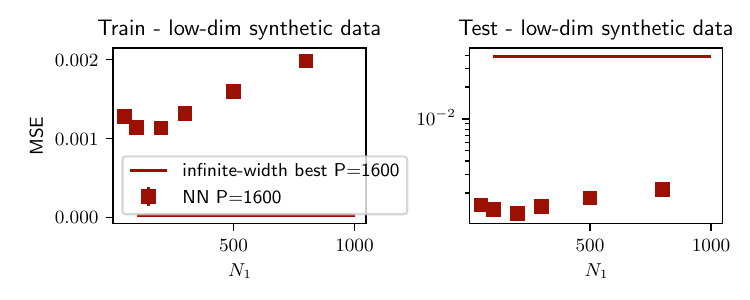}
    \caption{ \textbf{Bayesian training leads to overfitting with low-dimensional data.} MSE for train (left) and test (right) sets as a function of the hidden layer size $N_1$. The synthetic dataset used consists of $P = 1600$ examples with $N_0 = 12$ and $D$ scalar labels. In all panels, dots represent the best performance of a 1HL networks trained with Adam, solid lines represent the best infinite-width predictions. Even though the performances on the trainset are comparable between the NN and the infinite-width, the latter model overfits on the test set. The low-dimensional case where $N_0 \ll P, N_1$ falls outside the validity of both.
} 
\label{fig:4}
\end{figure}

We performed an experiment similar to the one in Fig. 3, using a synthetic dataset with $N_0 = 12$ and $D = 10$ scalar labels. In order to generate the synthetic dataset, we draw $N_0 = 2000$ samples of $D=20$ input variables, each uniformly distributed over the interval $[-1, 1]$. These inputs are then transformed into multiple output labels using a set of highly non-linear and heterogeneous functional forms. More concretely, the transformations are defined as: 
    \begin{equation*}
\begin{aligned}
y_1 &= \sin(2\pi x_1) + \log(x_2 + 1) + \epsilon_1 \quad & y_2 &= \cos(2\pi x_3) + \frac{x_4^2}{10} + \epsilon_2 \\[6pt]
y_3 &= \frac{\tan(2\pi x_5)}{5} + \sqrt{\max(0, x_6)} + \epsilon_3 \quad & y_4 &= \frac{x_7^3}{20} - \sin(2\pi x_8) + \epsilon_4 \\[6pt]
y_5 &= \frac{e^{x_9} - x_{10}}{20} + \epsilon_5 & &
\end{aligned}
\end{equation*}
while the remaining input dimensions are left unused in the output definitions, ensuring a scenario where only a subset of the input features is relevant to the prediction task.
In this setup, each output $y_i$ is perturbed by an additive Gaussian noise term $\epsilon_i$. Each $\epsilon_i$ is drawn independently from a normal distribution with zero mean and a variance proportional to the variance of the corresponding noiseless target, i.e. $\epsilon_i \sim \mathcal{N}(0, \sigma_i^2)$, where $\sigma_i^2 \propto \mathrm{Var}(y_i)$ with a chosen proportionality constant. This ensures that outputs with inherently larger variability receive correspondingly larger noise.
The dataset is then split into training and test sets, and both input and output values are normalized or scaled. This process yields a challenging multi-output regression dataset that shows various non-linear dependencies. A comparison between performances of infinite-width Gaussian Process and SOTA regression are shown in Fig \ref{fig:4}.

\end{document}